\documentclass[twocolumn,showpacs,preprintnumbers,amsmath,amssymb]{revtex4}
\usepackage{graphicx}

\begin{document}
\title{
%All-or-nothing
Oblivious transfer using quantum entanglement}
\author{Guang Ping He}
\affiliation{Department of Physics,
%\& Center of Theoretical and Computational Physics,
The University of Hong Kong, Pokfulam Road, Hong Kong, China}
\affiliation{
Department of Physics \& Advanced Research Center, Zhongshan University, Guangzhou\\
510275, China}
\author{Z. D. Wang}
%\email{zwang@hkucc.hku.hk}
\affiliation{Department of Physics \&
Center of Theoretical and Computational Physics, The University of
Hong Kong, Pokfulam Road, Hong Kong, China} \affiliation{National
Laboratory of Solid State Microstructures, Nanjing University,
Nanjing 210093, China}

\begin{abstract}
Based on quantum entanglement, an all-or-nothing oblivious
transfer protocol is proposed and is proven to be secure. The
distinct merit of the present protocol lies in that it is not
based on quantum bit commitment. More intriguingly, this OT
protocol
%is different from previous insecure ones and
does not belong to a class of protocols denied by the Lo's no-go
theorem of one-sided two-party secure computation, and thus its
security can be achieved.
%restricted by the no-go theorem of
\end{abstract}

\maketitle

\newpage

\section{Introduction}

The research on the oblivious transfer (OT) problem may be traced back to
more than twenty years ago\cite{Ra81,Wiesner}. Kilian\cite{Kilian} pointed
out later that OT is very important in two-party and multi-party protocols.
This is because in most symmetrical protocols, the participants always know
each others' data. If some participants are dishonest or try to get extra
information, the protocols become insecure. OT can create some secret
between the participants and break this symmetry. Thus it can be used to
implement multi-party protocols such as two-party secure computation\cite
{Crepeau95}. However, the security of classical cryptography usually has to
be based on some strong computational assumptions, such as the hardness of
factoring. If quantum computers become practical in the future, the validity
of these assumptions can be broken easily\cite{Shor}. Therefore significant
interests have been paid to look for quantum methods applicable to
cryptography to achieve better security\cite{BB84,Ekert}. Quantum oblivious
transfer (QOT) protocols were also proposed\cite{BBCS92}. But they are
secure only under the assumption that the participants cannot delay the
quantum measurement. To fix the problem, Cr\'{e}peau\cite{OT} proposed a QOT
protocol based on quantum bit commitment (QBC). It was further proven by Yao
\cite{Yao} that such a QOT is secure if QBC is secure. Nevertheless, it was
indicated later by Mayers, Lo and Chau that all the QBC protocols formerly
proposed\cite{BB84,BCJL93} are insecure. Furthermore, it was concluded that
unconditionally secure QBC scheme cannot be achieved in principle\cite
{Mayers,Lo}, which is referred to as the Mayers-Lo-Chau (MLC) no-go theorem
and is a serious drawback in quantum cryptography. According to the theorem,
all QBC based protocols are insecure, including quantum coin tossing\cite
{BB84,CT1,CT2,CT3} and quantum oblivious mutual identification\cite{OMI}.
Consequently, QOT based on QBC is insecure unless the participants are
restricted to individual measurements\cite{MS94}.

On the other hand, starting with QBC is not the only way to implement QOT.
Therefore, it is natural to ask whether we can design a QOT protocol with
stand-alone security. Although it was concluded independently that other
two-party quantum secure computations including QOT are not possible either
\cite{impossible1,impossible2}, the conclusion is essentially based on a
crucial point that the quantum state used in the two-party computation
protocols is the simultaneous eigenstate of different measurement operators,
which follows from two basic requirements in their definition of the
so-called ideal one-sided two-party secure computation\cite{impossible1}:
Alice helps Bob to compute a prescribed function $f(i,j)$\ in such a way
that, at the end of the protocol, (a) Bob learns $f(i,j)$\ unambiguously,
and (b) Alice learns nothing.
% which is used by the participant to determine the parameters he wants.
In this paper, a novel quantum OT protocol is proposed, which is neither
based on QBC nor satisfying rigorously the requirement (a); but it indeed
meets the rigorous security requirement of the OT definition. %Moreover,
%it was proven in classical cryptography that OT implies bit commitment.
%Though this equivalence may need more examination in the quantum level,
%our OT protocol is distinctly different from  previous  QBC
%protocols\cite{BCJL93} shown to be insecure by the MLC no-go
%theorem:  the cheater' strategy is assumed to execute the honest
%algorithm at the quantum level in the proof, implying that the
%honest participants never need to execute the honest algorithm at
%the quantum level; while this assumption is violated in our
%protocol to be detailed later.
Therefore, our OT protocol is a kind of two-party secure computation
%slightly
different from that defined by Lo in Ref.\cite{impossible1} and thus
%is not restricted by
evades the Lo's no-go theorem of the %ideal
one-sided two-party secure computation, allowing more quantum-cryptography
applications than thought possible previously. %The calculation of

In the next section, a new QOT protocol is elaborated in details. Then a
general proof of its unconditional security is presented in Sec. III.
Finally, the relationship between the protocol and the no-go theorems is
addressed.

\section{The scheme}

Although there are various types of OT, as a typical illustration,
%as a typical illustration,
%In the present work,
we here focus only on a basic type OT studied in Refs.\cite{Ra81,Yao}, which
is also called all-or-nothing OT. A sender Alice wants to transfer a secret
bit $b\in \{0,1\}$ to a receiver Bob. At the end of the protocol, either Bob
could learn the value of $b$ with the reliability (which means the
probability for Bob's output $b$ to be equal to Alice's input) $100\%$, or
he has zero knowledge on $b$. Each possibility should occur with the
probability $1/2$, and which one happens finally is out of their control.
Meanwhile, Alice should learn nothing about which event takes place.

Consider an ideal case without transmission error. Similar to the conjugate
coding\cite{Wiesner}, letting $\left| 0\right\rangle _{+}$\ and $\left|
1\right\rangle _{+}$ denote the two orthogonal states of a qubit, we can
define $\left| r\right\rangle _{\times }\equiv (\left| 0\right\rangle
_{+}+(-1)^{r}\left| 1\right\rangle _{+})/\sqrt{2}$ ($r=0,1$),
%and $\left| 1\right\rangle _{\times }\equiv
%(\left| 0\right\rangle _{+}-\left| 1\right\rangle _{+})/\sqrt{2}$,
the Bell states $\Phi ^{\pm }\equiv (\left| 0\right\rangle _{+}\left|
0\right\rangle _{+}\pm \left| 1\right\rangle _{+}\left| 1\right\rangle _{+})/%
\sqrt{2}$, and $\Psi ^{\pm }\equiv (\left| 0\right\rangle _{+}\left|
1\right\rangle _{+}\pm \left| 1\right\rangle _{+}\left| 0\right\rangle _{+})/%
\sqrt{2}$\ , where $+$ ($\times $) stands for the rectilinear (diagonal)
basis. \textit{The key idea of our protocol} is: Alice and Bob share many
sets of 4 qubits in an entangled state $\left| \psi \right\rangle $ (see Eq.(%
\ref{psai}) below). To each set, four two-value
parameters $q$, $r$, $c$, and $d$ are associated, where $\{q,r\}$ and $%
\{c,d\}$ correspond respectively to the state $\left| \psi
\right\rangle $ and the choice/measurement of individual
participant; the form of $\left| \psi \right\rangle $ designed by
us ensures that Alice cannot decode simultaneously any two of $q$,
$r$ and $d$, and Bob cannot decode $c$ and $q$ (or $r$)
simultaneously. Relying on appropriate verification and use of
state, a secure OT can be achieved.

 For easy readability, before presenting a complete version of our protocol,
 we first account for the details in several key procedures comprehensibly.

(i) \textit{Preparation of the states:}

Our protocol is based on the four-qubit entangled state with the following
form
\begin{eqnarray}
\left| \psi \right\rangle &=&\left| \psi _{A_{1}}\psi _{A_{2}}\psi
_{B_{1}}\psi _{B_{2}}\right\rangle  \nonumber \\
&=&(\left| 0\right\rangle _{+}\left| 0\right\rangle _{+}\left|
0\right\rangle _{+}\left| 0\right\rangle _{+}  \nonumber \\
&&+\left| 1\right\rangle _{+}\left| 1\right\rangle _{+}\left| 0\right\rangle
_{+}\left| 1\right\rangle _{+}  \nonumber \\
&&+\left| 0\right\rangle _{\times }\left| 0\right\rangle _{\times }\left|
1\right\rangle _{+}\left| 0\right\rangle _{+}  \nonumber \\
&&+\left| 1\right\rangle _{\times }\left| 1\right\rangle _{\times }\left|
1\right\rangle _{+}\left| 1\right\rangle _{+})/2.  \label{psai}
\end{eqnarray}
Bob prepares many sets of such states. For each set, he keeps systems $B_{1}$
and $B_{2}$\ and sends systems $A_{1}$ and $A_{2}$\ to Alice.

(ii) \textit{Alice inputting }$c$\textit{:}

In Alice's point of view,  Bob sends her any of the four two-qubit
states $\left| r\right\rangle _{q}\left| r\right\rangle _{q}$
($q\in \{+,\times \}$, $r\in \{0,1\}$) with the equal probability.
Now let us consider Alice's strategy to decode either $q$ or $r$.
In the Bell basis
\begin{equation}
C_{0}\equiv \{\Phi ^{+},\Phi ^{-},\Psi ^{+},\Psi ^{-}\},
\end{equation}
the four possible $\left| r\right\rangle _{q}\left| r\right\rangle
_{q}$ can be expressed as
\begin{eqnarray}
\left| 0\right\rangle _{+}\left| 0\right\rangle _{+} &=&(\Phi ^{+}+\Phi
^{-})/\sqrt{2},  \nonumber \\
\left| 1\right\rangle _{+}\left| 1\right\rangle _{+} &=&(\Phi ^{+}-\Phi
^{-})/\sqrt{2},  \nonumber \\
\left| 0\right\rangle _{\times }\left| 0\right\rangle _{\times } &=&(\Phi
^{+}+\Psi ^{+})/\sqrt{2},  \nonumber \\
\left| 1\right\rangle _{\times }\left| 1\right\rangle _{\times } &=&(\Phi
^{+}-\Psi ^{+})/\sqrt{2}.
\end{eqnarray}
If Alice measures systems $A_{1}$ and $A_{2}$ in the $C_{0}$ basis, she will
know that $q=+$ ($q=\times $) if the outcome is $\Phi ^{-}$\ ($\Psi ^{+}$).
While if the outcome is $\Phi ^{+}$, she will not know the value of $q$.
Since Eq.(\ref{psai}) can be rewritten as
\begin{eqnarray}
\left| \psi \right\rangle =\Phi ^{-}\left| 0\right\rangle _{+}\left|
1\right\rangle _{\times }/2 &&+\Psi ^{+}\left| 1\right\rangle _{+}\left|
1\right\rangle _{\times }/2  \nonumber \\
+\Phi ^{+}\left| 0\right\rangle _{\times }\left| 0\right\rangle _{\times }
&&/\sqrt{2},  \label{psaic0}
\end{eqnarray}
it can be seen that  the probability for Alice to  decode $q$
successfully is $1/2$.

On the other hand, defining the basis
\begin{equation}
C_{1}\equiv \{\left| 0\right\rangle _{\times }\left| 0\right\rangle
_{+},\left| 0\right\rangle _{\times }\left| 1\right\rangle _{+},\left|
1\right\rangle _{\times }\left| 0\right\rangle _{+},\left| 1\right\rangle
_{\times }\left| 1\right\rangle _{+}\},
\end{equation}
$\left| r\right\rangle _{q}\left| r\right\rangle _{q}$ can be expressed as
\begin{eqnarray}
\left| 0\right\rangle _{+}\left| 0\right\rangle _{+} &=&(\left|
0\right\rangle _{\times }\left| 0\right\rangle _{+}+\left| 1\right\rangle
_{\times }\left| 0\right\rangle _{+})/\sqrt{2},  \nonumber \\
\left| 1\right\rangle _{+}\left| 1\right\rangle _{+} &=&(\left|
0\right\rangle _{\times }\left| 1\right\rangle _{+}-\left| 1\right\rangle
_{\times }\left| 1\right\rangle _{+})/\sqrt{2},  \nonumber \\
\left| 0\right\rangle _{\times }\left| 0\right\rangle _{\times } &=&(\left|
0\right\rangle _{\times }\left| 0\right\rangle _{+}+\left| 0\right\rangle
_{\times }\left| 1\right\rangle _{+})/\sqrt{2},  \nonumber \\
\left| 1\right\rangle _{\times }\left| 1\right\rangle _{\times } &=&(\left|
1\right\rangle _{\times }\left| 0\right\rangle _{+}-\left| 1\right\rangle
_{\times }\left| 1\right\rangle _{+})/\sqrt{2}.
\end{eqnarray}
That is, if Alice measures them in the $C_{1}$ basis, she will
know that $r=0 $ ($r=1$) if the outcome is $\left| 0\right\rangle
_{\times }\left| 0\right\rangle _{+}$\ ($\left| 1\right\rangle
_{\times }\left| 1\right\rangle _{+}$), while she does not know
$r$ if the outcome is $\left| 0\right\rangle _{\times }\left|
1\right\rangle _{+}$ or $\left| 1\right\rangle _{\times }\left|
0\right\rangle _{+}$. Again, rewriting Eq.(\ref {psai}) as
\begin{eqnarray}
\left| \psi \right\rangle  &=&\left| 0\right\rangle _{\times }\left|
0\right\rangle _{+}\left| 0\right\rangle _{\times }\left| 0\right\rangle
_{+}/2-\left| 1\right\rangle _{\times }\left| 1\right\rangle _{+}\left|
0\right\rangle _{\times }\left| 1\right\rangle _{+}/2  \nonumber \\
&&+\left| 0\right\rangle _{\times }\left| 1\right\rangle _{+}\Psi
^{+}/2+\left| 1\right\rangle _{\times }\left| 0\right\rangle _{+}\ \Phi
^{+}/2,  \label{psaic1}
\end{eqnarray}
we see that the  probability for Alice to decode $r$ successfully
is also $1/2$.
%The above results of Alice's measurements are
%schematically illustrated in Fig.1.
Also, since the bases $C_{0}$
and $C_{1}$ are not commutable, Alice cannot decode the values of
$q$ and $r$ simultaneously (A rigorous proof will be provided in
the next section).

In our protocol, Alice should randomly picks a different bit $c\in \{0,1\}$
for each set of $\left| \psi \right\rangle $ at this stage. If $c=0$ ($c=1$%
), she tries to decode $q$ ($r$) by measuring her share of the set in the $%
C_{0}$\ ($C_{1}$) basis. After she measures all $\left| \psi \right\rangle $%
, she will decode either $q$ or $r$ successfully for about half of
these sets, while she fails to decode anything for the other half.
She tells Bob to discard the half which she failed to decode,
while keeps the rest sets of $\left| \psi \right\rangle $\ in the
following steps.

Bob can verify whether Alice has input $c$ and finished her
measurement by  picking randomly some $\left| \psi \right\rangle
$\ from the remaining half, and asking Alice to announce either
$q$ or $r$, depending on what she decoded. To find out the correct
value of $q$ or $r$, as can be seen from Eq.(\ref{psai}), Bob can
simply measures systems $B_{1}$ and $B_{2}$ of the picked $\left|
\psi \right\rangle $\ in the basis
\begin{equation}
D_{0}\equiv \{\left| 0\right\rangle _{+}\left| 0\right\rangle _{+},\left|
0\right\rangle _{+}\left| 1\right\rangle _{+},\left| 1\right\rangle
_{+}\left| 0\right\rangle _{+},\left| 1\right\rangle _{+}\left|
1\right\rangle _{+}\}.  \label{D0}
\end{equation}
Then he learns which $\left| r\right\rangle _{q}\left|
r\right\rangle _{q}$ systems $A_{1}$ and $A_{2}$ can collapse to.
If Alice has delayed her measurement or adopted any other
measurement which is less efficient than the above strategies on
decode $q$ or $r$ with certainty, she cannot always announce $q$
or $r$ correctly, or she has to discard more than half of the
sets. Therefore a dishonest Alice will inevitably be caught as the
increase of the number of $\left| \psi \right\rangle $\ picked for
the verification.

Nevertheless,  to pass the verification, Alice needs not to
perform complete measurement in the $C_{0}$\ ($C_{1}$) basis. She
can simply try to project systems $A_{1}$ and $A_{2}$ to the
subspace supported by $\{\Phi ^{-},\Psi ^{+}\}$\ ($\{\left|
0\right\rangle _{\times }\left| 0\right\rangle _{+},\left|
1\right\rangle _{\times }\left| 1\right\rangle _{+}\}$). If the
projection fails, she tells Bob to discard the corresponding
$\left| \psi \right\rangle $. While if the projection is
successful, she keeps systems $A_{1}$ and $A_{2}$ entangled with
$B_{1}$ and
$B_{2}$ without collapsing them into a pure state $\Phi ^{-}$ or $\Psi ^{+}$%
\ ($\left| 0\right\rangle _{\times }\left| 0\right\rangle _{+}$ or $\left|
1\right\rangle _{\times }\left| 1\right\rangle _{+}$). She finishes the
complete measurement to make them collapse only when the corresponding $%
\left| \psi \right\rangle $\ is picked for the verification. Therefore in
general, the state of the remaining undiscarded and unverified sets of $%
\left| \psi \right\rangle $\ is either
\begin{equation}
\left| \psi ^{(0)}\right\rangle =\Phi ^{-}\left| 0\right\rangle
_{+}\left| 1\right\rangle _{\times }/\sqrt{2}+\Psi ^{+}\left|
1\right\rangle _{+}\left| 1\right\rangle _{\times }/\sqrt{2}
\label{psai0}
\end{equation}
if $c=0$, or
\begin{equation}
\left| \psi ^{(1)}\right\rangle =\left| 0\right\rangle _{\times
}\left| 0\right\rangle _{+}\left| 0\right\rangle _{\times }\left|
0\right\rangle _{+}/\sqrt{2}-\left| 1\right\rangle _{\times
}\left| 1\right\rangle _{+}\left| 0\right\rangle _{\times }\left|
1\right\rangle _{+}/\sqrt{2} \label{psai1}
\end{equation}
if $c=1$. After the verification, Alice and Bob keep these $\left| \psi
\right\rangle $\ and proceed.

(iii) \textit{Bob inputting }$d$\textit{:}

Since the state of systems $B_{1}$ and $B_{2}$ are different in Eqs.(\ref
{psai0}) and (\ref{psai1}), Bob can learn Alice's choice of $c$ or her
outcome $s$ with a certain probability. Here Alice's outcome $s$ is defined
as
\begin{equation}
s\equiv \left\{
\begin{array}{c}
Q,\qquad (c=0), \\
r,\qquad (c=1),
\end{array}
\right.   \label{s}
\end{equation}
where $Q=0,1$\ for $q=+,\times $. From Eq.(\ref{psai}) we can see that if
Bob measures systems $B_{1}$ and $B_{2}$ in the $D_{0}$ basis defined in Eq.(%
\ref{D0}) and the outcome is $\left| 0\right\rangle _{+}\left|
0\right\rangle _{+}$ (or $\left| 1\right\rangle _{+}\left| 1\right\rangle
_{+}$), he will know that systems $A_{1}$ and $A_{2}$ can only collapse to
the state $\left| 0\right\rangle _{+}\left| 0\right\rangle _{+}$ (or $\left|
1\right\rangle _{\times }\left| 1\right\rangle _{\times }$). These two
states have the common feature $Q=r$. Thus Bob knows that $s=0$ ($s=1$)
despite he does not know $c$.

Note that at this stage, $\left| \psi \right\rangle $ already collapsed to $%
\left| \psi ^{(0)}\right\rangle $\ or $\left| \psi ^{(1)}\right\rangle $\ by
Alice's measurement. With the $D_{0}$ basis, they can be expressed as
\begin{eqnarray}
\left| \psi ^{(0)}\right\rangle &=&[\Phi ^{-}(\left| 0\right\rangle
_{+}\left| 0\right\rangle _{+}-\left| 0\right\rangle _{+}\left|
1\right\rangle _{+})  \nonumber \\
&&+\Psi ^{+}(\left| 1\right\rangle _{+}\left| 0\right\rangle _{+}-\left|
1\right\rangle _{+}\left| 1\right\rangle _{+})]/2,  \label{psai0d0}
\end{eqnarray}
and
\begin{eqnarray}
\left| \psi ^{(1)}\right\rangle &=&[\left| 0\right\rangle _{\times }\left|
0\right\rangle _{+}(\left| 0\right\rangle _{+}\left| 0\right\rangle
_{+}+\left| 1\right\rangle _{+}\left| 0\right\rangle _{+})  \nonumber \\
&&-\left| 1\right\rangle _{\times }\left| 1\right\rangle _{+}(\left|
0\right\rangle _{+}\left| 1\right\rangle _{+}+\left| 1\right\rangle
_{+}\left| 1\right\rangle _{+})]/2.  \label{psai1d0}
\end{eqnarray}
Thus the  probability for Bob to decode $s$ successfully is $1/2$.

On the other hand, defining the basis
\begin{equation}
D_{1}\equiv \{\left| 0\right\rangle _{\times }\left| 0\right\rangle _{\times
},\left| 0\right\rangle _{\times }\left| 1\right\rangle _{\times },\left|
1\right\rangle _{\times }\left| 0\right\rangle _{\times },\left|
1\right\rangle _{\times }\left| 1\right\rangle _{\times }\},
\end{equation}
$\left| \psi ^{(0)}\right\rangle $\ or $\left| \psi ^{(1)}\right\rangle $\
can be expressed as
\begin{equation}
\left| \psi ^{(0)}\right\rangle =(\Phi ^{-}-\Psi ^{+})\left| 1\right\rangle
_{\times }\left| 1\right\rangle _{\times }/2+(\Phi ^{-}+\Psi ^{+})\left|
0\right\rangle _{\times }\left| 1\right\rangle _{\times }/2,  \label{psai0d1}
\end{equation}
and
\begin{eqnarray}
\left| \psi ^{(1)}\right\rangle  &=&(\left| 0\right\rangle _{\times }\left|
0\right\rangle _{+}-\left| 1\right\rangle _{\times }\left| 1\right\rangle
_{+})\left| 0\right\rangle _{\times }\left| 0\right\rangle _{\times }/2
\nonumber \\
&&+(\left| 0\right\rangle _{\times }\left| 0\right\rangle _{+}+\left|
1\right\rangle _{\times }\left| 1\right\rangle _{+})\left| 0\right\rangle
_{\times }\left| 1\right\rangle _{\times }/2.  \label{psai1d1}
\end{eqnarray}
If Bob measures systems $B_{1}$ and $B_{2}$\ in the $D_{1}$ basis,
he will know that $c=0$ ($c=1$) if the outcome is $\left|
1\right\rangle _{\times }\left| 1\right\rangle _{\times }$\
($\left| 0\right\rangle _{\times }\left| 0\right\rangle _{\times
}$), while he does not know $c$ if the outcome is $\left|
0\right\rangle _{\times }\left| 1\right\rangle _{\times }$. The
probability for him to decode $c$ successfully is also $1/2$.
%The results of Bob's measurement are schematically summarized in Fig.2.
Again, Bob cannot decode the values of $s$ and $c$
simultaneously since the bases $D_{0} $ and $D_{1}$ are not
commutable.

In the protocol, Bob randomly picks a different bit $d\in \{0,1\}$ for each
remaining set of $\left| \psi \right\rangle $, where $d=0$ should occur with
the probability $p=2/3$ (we will see later why this value is chosen). If $d=0
$ ($d=1$), he tries to decode $s$ ($c$) by measuring his share of the set in
the $D_{0}$\ ($D_{1}$) basis. After he measures all sets of $\left| \psi
\right\rangle $, he will decode either $s$ or $c$ successfully for about
half of those sets, while he fails to decode anything for the other half. He
tells Alice to discard the half which he failed to decode, while keeping the
rest $\left| \psi \right\rangle $\ for the following steps.

Similar to (ii), Alice can verify whether Bob has input $d$ and
finished his measurement honestly by  picking randomly some
$\left| \psi \right\rangle $\ from the remaining half, and asking
Bob to announce either $s$ or $c$, depending on what he decoded.
She should also check whether Bob has indeed input $d=0$ with the
required probability $p=2/3$, and whether the number of discarded
$\left| \psi \right\rangle $\ is about a half.

Also, Bob needs not to perform a complete measurement in the $D_{0}$\ ($D_{1}$%
) basis\ to pass the verification. If he has chosen $d=0$ ($d=1$), he simply
tries to project systems $B_{1}$ and $B_{2}$ to the subspace supported by $%
\{\left| 0\right\rangle _{+}\left| 0\right\rangle _{+},\left|
1\right\rangle _{+}\left| 1\right\rangle _{+}\}$\ ($\{\left|
0\right\rangle _{\times }\left| 0\right\rangle _{\times },\left|
1\right\rangle _{\times }\left| 1\right\rangle _{\times }\}$), and
discards $\left| \psi \right\rangle $ if the projection fails. He
finishes the complete measurement to make the undiscarded $\left|
\psi \right\rangle $ collapse only when it\ is picked for the
verification. Therefore after the verification, the state of the
remaining  unverified $\left| \psi \right\rangle $ is
\begin{equation}
\left| \psi ^{(00)}\right\rangle =(\Phi ^{-}\left| 0\right\rangle
_{+}\left| 0\right\rangle _{+}-\Psi ^{+}\left| 1\right\rangle
_{+}\left| 1\right\rangle _{+})/\sqrt{2}
\end{equation}
if $c=0$ and $d=0$, or
\begin{equation}
\left| \psi ^{(10)}\right\rangle =(\left| 0\right\rangle _{\times
}\left| 0\right\rangle _{+}\left| 0\right\rangle _{+}\left|
0\right\rangle _{+}-\left| 1\right\rangle _{\times }\left|
1\right\rangle _{+}\left| 1\right\rangle _{+}\left| 1\right\rangle
_{+})/\sqrt{2}
\end{equation}
if $c=1$ and $d=0$, or
\begin{equation}
\left| \psi ^{(01)}\right\rangle =(\Phi ^{-}-\Psi ^{+})\left| 1\right\rangle
_{\times }\left| 1\right\rangle _{\times }/\sqrt{2},  \label{psai01}
\end{equation}
if $c=0$ and $d=1$, or
\begin{equation}
\left| \psi ^{(11)}\right\rangle =(\left| 0\right\rangle _{\times }\left|
0\right\rangle _{+}-\left| 1\right\rangle _{\times }\left| 1\right\rangle
_{+})\left| 0\right\rangle _{\times }\left| 0\right\rangle _{\times }/\sqrt{2%
},  \label{psai11}
\end{equation}
if $c=1$ and $d=1$.

Before using these states for the OT, Bob must prevent Alice from knowing
his choice of $d$ for each of them. It can be accomplished with the
following method. $\left| \psi ^{(00)}\right\rangle $\ and $\left| \psi
^{(10)}\right\rangle $\ can be rewritten as
\begin{equation}
\left| \psi ^{(00)}\right\rangle =(\Phi ^{-}+\Psi ^{+})\Phi ^{-}/2+(\Phi
^{-}-\Psi ^{+})\Phi ^{+}/2,  \label{psai00}
\end{equation}
and
\begin{eqnarray}
\left| \psi ^{(10)}\right\rangle  &=&(\left| 0\right\rangle _{\times }\left|
0\right\rangle _{+}+\left| 1\right\rangle _{\times }\left| 1\right\rangle
_{+})\Phi ^{-}/2  \nonumber \\
&&+(\left| 0\right\rangle _{\times }\left| 0\right\rangle _{+}-\left|
1\right\rangle _{\times }\left| 1\right\rangle _{+})\Phi ^{+}/2.
\label{psai10}
\end{eqnarray}
If Bob has chosen $d=0$, he tries to project systems $B_{1}$ and $B_{2}$ to
the state $\Phi ^{+}$, and tells Alice to discard the corresponding $\left|
\psi \right\rangle $ if the projection fails. Then the remaining $\left|
\psi ^{(00)}\right\rangle $\ and $\left| \psi ^{(10)}\right\rangle $\
collapse to
\begin{equation}
\left| \psi ^{(00a)}\right\rangle =(\Phi ^{-}-\Psi ^{+})\Phi ^{+}/\sqrt{2},
\label{psai00a}
\end{equation}
and
\begin{equation}
\left| \psi ^{(10a)}\right\rangle =(\left| 0\right\rangle _{\times }\left|
0\right\rangle _{+}-\left| 1\right\rangle _{\times }\left| 1\right\rangle
_{+})\Phi ^{+}/\sqrt{2}.  \label{psai10a}
\end{equation}
We can see that the states of systems $A_{1}$ and $A_{2}$ of
$\left| \psi ^{(00a)}\right\rangle $ and $\left| \psi
^{(10a)}\right\rangle $\ (which are corresponding to $d=0$) are
exactly the same as those of $\left| \psi ^{(01)}\right\rangle $
and $\left| \psi ^{(11)}\right\rangle $ (corresponding to $d=1$)
respectively. Therefore Alice can by no means distinguish them
apart so she cannot know $d$ from the remaining $\left| \psi
\right\rangle $.

Now let us explain why Bob should choose $d=0$ with the probability $p=2/3$.
Eqs.(\ref{psai00}) and (\ref{psai10}) show that half of the $\left| \psi
\right\rangle $ corresponding to $d=0$ will be further discarded when
collapsing $\left| \psi ^{(00)}\right\rangle $\ and $\left| \psi
^{(10)}\right\rangle $\ to $\left| \psi ^{(00a)}\right\rangle $ and $\left|
\psi ^{(10a)}\right\rangle $. Meanwhile, no $\left| \psi \right\rangle $
corresponding to $d=1$ will be discarded after Alice has verified Bob's
action. Therefore among all the remaining $\left| \psi \right\rangle $, $d=0$%
\ and $d=1$\ will occur with the equal probability $1/2$, which
will be useful below.

(iv) \textit{Completing the OT:}

At this stage, for any remaining $\left| \psi \right\rangle $, Alice knows
her own choice $c$ but not Bob's choice $d$, while Bob has chosen $d=0$\
(i.e. he does not knows $c$) and $d=1$\ (he knows $c$) with the equal
probability $1/2$. Thus Alice can randomly pick any one of the remaining $%
\left| \psi \right\rangle $, and use $c$\ to encode the bit $b$
she wants to transfer. If by chance Bob knows $c$ for this chosen
$\left| \psi \right\rangle $,\ he can decode $b$ successfully.
Else he knows nothing about $b$. Because the two results will
occur with the equal probability $1/2$, the goal of OT is
accomplished.

The above procedure is summarized as the protocol below, with the
corresponding schematic flow chart being illustrated in Fig.1.

\bigskip

\textit{Protocol OT}

(1) \textit{Preparation of the states:} Bob prepares $n$\ sets of $\left|
\psi \right\rangle $ as described in Eq.(\ref{psai}). He keeps systems $%
B_{1} $ and $B_{2}$\ of each $\left| \psi \right\rangle $ and sends systems $%
A_{1}$ and $A_{2}$\ to Alice;

(2) \textit{Alice inputting }$c$\textit{:}

\qquad (2-1) For each $\left| \psi \right\rangle $, Alice views the state of
systems $A_{1}$ and $A_{2}$\ as $\left| r\right\rangle _{q}\left|
r\right\rangle _{q}$, and she randomly picks $c\in \{0,1\}$. If $c=0$, She
tries to decode $q$ by projecting the two qubits into $\Phi ^{-}$ and $\Psi
^{+}$, and she sets $q=+$ ($q=\times $) if the outcome is $\Phi ^{-}$ ($\Psi
^{+}$). Else if $c=1$, Alice tries to decode $r$ by projecting the two
qubits into $\left| 0\right\rangle _{\times }\left| 0\right\rangle _{+}$ and
$\left| 1\right\rangle _{\times }\left| 1\right\rangle _{+}$, and \ she sets
$r=0$ ($r=1$) if the outcome is $\left| 0\right\rangle _{\times }\left|
0\right\rangle _{+}$ ($\left| 1\right\rangle _{\times }\left| 1\right\rangle
_{+}$);

\qquad (2-2) If the projection in (2-1) fails, Alice tells Bob to discard
the corresponding $\left| \psi \right\rangle $;

(3) \textit{Verification 1:}

\qquad (3-1) If the number of the remaining $\left| \psi \right\rangle $ is $%
n^{\prime }\backsim n/2$ they continue\cite{note}, else they abort the
procedure;

\qquad (3-2) Bob randomly picks some of the remaining $\left| \psi
\right\rangle $ and asks Alice to announce either their $q$ or $r$ depending
on the value of $c$. To check Alice's announcement, Bob measures $\psi
_{B_{1}}\psi _{B_{2}}$\ in the $D_{0}$\ basis,\ and uses the result to
calculate $q$, $r$ that corresponds to $\psi _{A_{1}}\psi _{A_{2}}$;

\qquad (3-3) Alice randomly picks some other remaining $\left| \psi
\right\rangle $\ and asks Bob to announce both $q$ and $r$. Bob performs the
same measurement in (3-2) to obtain $q$, $r$ to announce;

\qquad (3-4) If $\{$no conflicting results were found by both participants$%
\} $ $AND$ $\{$the probabilities for $\left| r\right\rangle _{q}\left|
r\right\rangle _{q}=\left| 0\right\rangle _{+}\left| 0\right\rangle _{+}$, $%
\left| r\right\rangle _{q}\left| r\right\rangle _{q}=\left| 1\right\rangle
_{+}\left| 1\right\rangle _{+}$, $\left| r\right\rangle _{q}\left|
r\right\rangle _{q}=\left| 0\right\rangle _{\times }\left| 0\right\rangle
_{\times }$ and $\left| r\right\rangle _{q}\left| r\right\rangle _{q}=\left|
1\right\rangle _{\times }\left| 1\right\rangle _{\times }$ to occur are
approximately the same$\}$, they keep the remaining undiscarded and
unverified $\left| \psi \right\rangle $ and continue;

(4) \textit{Bob inputting }$d$\textit{:}

\qquad (4-1) For each of the remaining $m$ sets of $\left| \psi
\right\rangle $, Bob picks $d=0$ with the probability $p=2/3$ and $d=1$ with
the probability $(1-p)=1/3$. If $d=0$, he tries to decode $s$ (defined as
Eq.(\ref{s})) by projecting $\psi _{B_{1}}\psi _{B_{2}}$ into the subspace
supported by $\{\left| 0\right\rangle _{+}\left| 0\right\rangle _{+},\left|
1\right\rangle _{+}\left| 1\right\rangle _{+}\}$. Else if $d=1$, Bob tries
to decode $c$\ by projecting $\psi _{B_{1}}\psi _{B_{2}}$ into $\left|
1\right\rangle _{\times }\left| 1\right\rangle _{\times }$ and $\left|
0\right\rangle _{\times }\left| 0\right\rangle _{\times }$. If the outcome
is $\left| 1\right\rangle _{\times }\left| 1\right\rangle _{\times }$ ($%
\left| 0\right\rangle _{\times }\left| 0\right\rangle _{\times }$), he knows
that Alice has chosen $c=0$ ($c=1$);

\qquad (4-2) If the projection in (4-1) fails, Bob tells Alice to discard
the corresponding $\left| \psi \right\rangle $;

(5) \textit{Verification 2:}

\qquad (5-1) If the number of the remaining $\left| \psi \right\rangle $\ is
about $m/2$ they continue; else they abort the procedure;

\qquad (5-2) Alice randomly picks some of the remaining $\left| \psi
\right\rangle $ and asks Bob to announce either $c$ or $s$ depending on the
value of $d$. Note that if $d=0$, Bob needs to complete the measurement on $%
\psi _{B_{1}}\psi _{B_{2}}$\ in the basis $\{\left| 0\right\rangle
_{+}\left| 0\right\rangle _{+},\left| 1\right\rangle _{+}\left|
1\right\rangle _{+}\}$,\ and he announces $s=0$ ($s=1$) if the outcome is $%
\left| 0\right\rangle _{+}\left| 0\right\rangle _{+}$\ ($\left|
1\right\rangle _{+}\left| 1\right\rangle _{+}$);

\qquad (5-3) If $\{$no conflicting results were found$\}$ $AND$ $\{d=0$
occurs with the probability $2/3\}$, they keep the remaining undiscarded and
unverified $\left| \psi \right\rangle $ and continue;

(6) \textit{Bob preventing Alice from knowing }$d$\textit{:} For each
remaining $\left| \psi \right\rangle $\ which Bob has chosen $d=0$, he tries
to project $\psi _{B_{1}}\psi _{B_{2}}$\ into the state $\Phi ^{+}$, and
tells Alice to discard the corresponding $\left| \psi \right\rangle $ if the
projection fails;

(7) \textit{OT part:}

\qquad (7-1) Alice randomly picks one of the remaining $\left| \psi
\right\rangle $ and tells Bob $b^{\prime }=b\oplus c$;

\qquad (7-2) If Bob has chosen $d=1$ for this $\left| \psi \right\rangle $
he calculates $b=b^{\prime }\oplus c$. Else he knows that he fails to get $b$%
.

\begin{figure}[b]
\includegraphics[width=8.5cm]{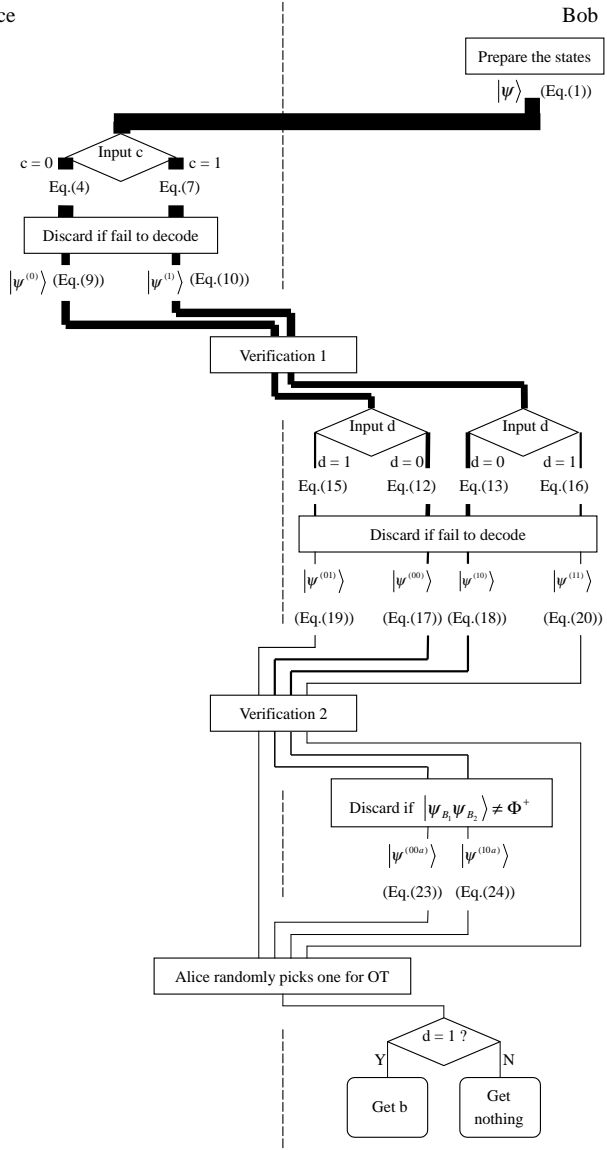}
\caption{\label{Figure} A schematic flow chart of Protocol OT. The
boxes on the left (right) represent the local operations on
Alice's (Bob's) side, while the middle ones are those requiring
collaboration of them. The width of the lines denote qualitatively
 the number of the states.}
\end{figure}

\section{Proof of security}

We now prove generally that the protocol is secure against any cheating
strategy in three steps: (I) the form of $\left| \psi \right\rangle $ limits
the knowledge of Alice and Bob; (II) the verifications limit both
participants' behaviors to honest ones; and (III) if Bob does not prepare $%
\left| \psi \right\rangle $ honestly, his knowledge on the data will be even
worse.

(I) \textit{For the state }$\left| \psi \right\rangle $\textit{\ given in
Eq.(\ref{psai}), Alice cannot learn }$q$\textit{, }$r$\textit{\ and }$d$%
\textit{\ simultaneously with the reliability }$100\%$\textit{, and Bob
cannot learn }$s$ \textit{and} $c$\ \textit{simultaneously\ with the
reliability }$100\%$\textbf{.}

\textit{Proof:} Let $\rho _{0}$ ($\rho _{1}$) denote the reduced density
matrix of the quantum state on Alice's side corresponding to $q=q_{0}$ and $%
r=r_{0}$ ($q\neq q_{0}$ and $r\neq r_{0}$). %Let $\rho _{0}$
%($\rho _{1}$) denote the density matrix of the quantum state on Alice's side
%whose $q_{i}$ and $r_{i}$ have certain values (whose $q_{i}$ and $r_{i}$ do
%not have these certain values).
To make sure that $q=q_{0}$ and $r=r_{0}$ simultaneously, Alice needs to
distinguish $\rho _{0}$ from $\rho _{1}$. It can be proven that the optimal
strategy for her to identify $\rho _{0}$ with the reliability $100\%$ is to
measure the states in the basis in which $\rho _{1}$ is diagonalized.
Supposing that $\rho _{0}$ and $\rho _{1}$ are expressed in this basis with $%
\rho (k,l)$ denoting the element of the matrix $\rho $, the maximum
probability for identifying $\rho _{0}$ is
\begin{equation}
p_{0\max }=\sum_{k\in \{k|\rho _{1}(k,k)=0\}}\rho _{0}(k,k).
\end{equation}
When $\left| \psi \right\rangle $ takes the form as specified in Eq.(\ref
{psai}), it is shown that $\{k|\rho _{1}(k,k)=0\}=\phi $ (the empty set)
regardless of the values of $q$ and $r$. Therefore $p_{0\max }=0$, which
means that Alice can never learn the exact values of $q$ and $r$
simultaneously with the reliability $100\%$. Similarly, it can also be
proven that Bob cannot learn $s$ and $c$\ simultaneously\ with the
reliability $100\%$.

As for $d$, by comparing Eqs.(\ref{psai01}) with (\ref{psai00a}) and Eqs.(%
\ref{psai11}) with (\ref{psai10a}) respectively, we can see that after the
step (6), the final states of systems $A_{1}$ and $A_{2}$ are exactly the
same regardless Bob's choice of $d$. Therefore Alice cannot learn $d$ as
long as the protocol can indeed force the participants to perform the honest
measurement. This leads us to the next point of the proof.

(II) \textit{For the state }$\left| \psi \right\rangle $,\textit{\ the steps
(3-1) and (3-2) can force Alice to measure the states honestly in the step
(2), and the step (5) can force Bob to measure the states honestly in the
step (4)}.

\textit{Proof:} Consider Alice's cheating first. Suppose that in the step
(3-2), there are totally $\delta n^{\prime }$ sets of $\left| \psi
\right\rangle $ which have not been measured by Alice honestly. Instead, she
applies a minimal-error measurement or even delays the measurement. Then she
does not know their $q$ or $r$ with the reliability $100\%$, but only with a
reliability being not larger than $\varepsilon $. As Bob picks randomly many
$\left| \psi \right\rangle $ to check if Alice knows $q$ or $r$, the
probability for Alice to pass the test is $\varepsilon ^{O(\delta n)}$.
Meanwhile, since only one $\left| \psi \right\rangle $ is randomly picked
for the OT at the final stage, the probability for these $\delta $ sets of $%
\left| \psi \right\rangle $ to be picked is not greater than $%
\sum\nolimits_{i=1}^{\min (\delta n^{\prime },m)}C_{\delta n^{\prime
}}^{i}C_{n^{\prime }-i}^{m-i}/C_{n^{\prime }}^{m}$. The order of magnitude
of this probability is $O(\delta )$ as long as $n>>m$. Therefore the total
probability for Alice to cheat successfully is bounded by $O(\delta
)\varepsilon ^{O(\delta n)}$, which can be made arbitrarily small as $%
n\rightarrow \infty $.

Thus Alice cannot use the minimal-error measurement, but has to use the
measurements which always decode $q$ or $r$ with the reliability $100\%$.
Here it is shown that the honest measurement is the optimal one among all
these measurements. Using the method described in (I), let $\rho _{0}$ and $%
\rho _{1}$ be the density matrices for $q=+$ and $\times $ respectively
(being independent of $r$). In the Bell basis, both $\rho _{0}$ and $\rho
_{1}$ are diagonalized. The maximum probabilities for Alice to identify them
are the same: $p_{0\max }=p_{1\max }=1/2$, which can be reached
simultaneously in the same measurement. Thus the maximum probability for
Alice to decode $q$ with the reliability $100\%$ is $p=(p_{0\max }+p_{1\max
})/2=1/2$. And the operation in the step (2) is just the strategy that can
reach this maximum. The calculation of the maximum probability for Alice to
decode $r$ successfully is a little bit more complicated. In this case, $%
\rho _{0}$ (for $r=0$) and $\rho _{1}$ (for $r=1$) cannot be diagonalized
simultaneously, and the maximum probability $p<(p_{0\max }+p_{1\max })/2$.
But we can see that when Alice chooses to decode $r$ in the step (2), if the
corresponding projection succeeds, she immediately gets $1$ bit of
information; while the projection fails, $\rho _{0}$ and $\rho _{1}$
collapse to the same density matrix, i.e., the upper bound of the average
information that can be gained from the resultant final states is zero. This
fact implies that Alice had already drawn as much information as possible
from the states she received. Therefore when $r=0$ and $r=1$ occur with the
same probability, the strategy in that step is exactly the optimal one for
her to get $r$ with the reliability $100\%$. The maximum probability of this
procedure is also $p=1/2$. Namely, Alice cannot decode $q$ or $r$\
unambiguously with a probability higher than that of the honest measurement.

As a result, if Alice makes her measurement without using the correct method
in the step (2) or even delays her measurement until Bob announces which
sets of $\left| \psi \right\rangle $ picked for the verification in the step
(3-2), either she cannot reach the maximum efficiency such that she has to
discard more data than what is allowed in the step (3-1), or there will
inevitably be some undiscarded $q$ or $r$ whose reliability is only $%
\varepsilon <100\%$. She cannot pass the test with a nontrivial probability,
because in the step (3-2) it is no longer allowed to discard the data that
she fails to decode. For this reason, Alice has to follow the protocol
honestly.

Repeating the above procedure, we can obtain the similar result for the case
in which Bob applies the minimal measurement or other dishonest
measurements. Bob has to choose $d=0$ and $d=1$\ with the specified ratio
and use the method in the step (4) to measure all $\psi _{B_{1}}\psi
_{B_{2}} $. Else he will only have a probability $O(\delta )\varepsilon
^{O(\delta m)} $ to cheat without being caught.

From the above (I) and (II), we can see that the goal of OT can be achieved,
as long as the initial state $\left| \psi \right\rangle $\ takes the
specific form given in Eq.(\ref{psai}). This allows us to proceed to the
last but not the least part of the proof.

(III) \textit{Steps (3-3) and (5) are able to force Bob to prepare the
states honestly.}

\textit{Proof:} The step (5) requires Bob to show that he has indeed input $%
d $ for all the remaining $\left| \psi \right\rangle $ (i.e., he already got
$c $ or $s$\ with the reliability $100\%$), while only about $m/2$ sets are
allowed to be discarded. Therefore for the same reason in (II), in the step
(4) the probability for Bob to get $c$ with the reliability $100\%$ should
reach $1/2$. We shall prove that, if Bob does not prepare the initial states
honestly, this probability will drop, or he will not pass the test in the
step (3-3).

Let us first study what constrain will be put on the initial states by the
step (3-3). There may exist many cheating strategies for Bob. But they can
all be described by the following model. Bob sends Alice a quantum system $%
\alpha $ which is entangled with another system $\beta $. He performs any
POVMs\cite{qi73} on $\beta $\ to get as much information as he can. A
general form of the entangled system $\alpha \otimes \beta $ is
\begin{equation}
\left| \psi \right\rangle =\sum_{k}f_{k}\left| \alpha _{k}\right\rangle
\left| \beta _{k}\right\rangle .
\end{equation}
Alice can check the partial density matrix $\rho _{\alpha
}=\sum_{k}\left\langle \beta _{k}\right. \left| \psi \right\rangle
\left\langle \psi \right| \left. \beta _{k}\right\rangle $\ of system $%
\alpha $ with her measurement in the step (3-3). Therefore to make Alice
believe that he is honest, Bob has to prepare $\alpha \otimes \beta $ in
such a way that $\rho _{\alpha }$ is much the same as that of $\psi
_{A_{1}}\psi _{A_{2}}$ in the honest protocol. Thus for each single set of $%
\left| \psi \right\rangle $, the state of such a system can be expanded as
\cite{qi73}
\begin{equation}
\left| \psi \right\rangle =\sum_{r\in \{0,1\},q\in \{+,\times
\}}f_{r,q}\left| r\right\rangle _{q}\left| r\right\rangle _{q}\left|
B_{r,q}\right\rangle .
\end{equation}
Bob sends Alice the first two qubits, and keeps the last part on his side as
$\beta $. Generally $\beta $\ can include any systems at Bob's side and the
environment, and even the systems $A_{1}$ and $A_{2}$ from other sets of $%
\left| \psi \right\rangle $ at Alice's side. But Bob does not know
beforehand which $\left| \psi \right\rangle $ will be picked for the test in
the step (3-3). So he needs to prepare $\beta $\ with the following
property: once the corresponding $\left| \psi \right\rangle $ is picked, he
can always measure $\beta $\ and get $q$, $r$ unmistakably. Thus $\beta $\
has to contain the systems on Bob's side only, and all the states $\left|
B_{r,q}\right\rangle $ with different $q$, $r$\ need to be orthogonal to
each other.

We now evaluate the amount of information on $c$ that Bob can obtain with
such a state. Suppose that $\left| \psi \right\rangle $ eventually survives
through the step (3). This state can be expressed as
\begin{eqnarray}
\left| \psi \right\rangle &=&(\sum_{r,q}f_{r,q}\Phi ^{+}\left|
B_{r,q}\right\rangle +\sum_{r}(-1)^{r}f_{r,+}\Phi ^{-}\left|
B_{r,+}\right\rangle  \nonumber \\
&&+\sum_{r}(-1)^{r}f_{r,\times }\Psi ^{+}\left| B_{r,\times }\right\rangle )/%
\sqrt{2}.
\end{eqnarray}
Since Alice already included this $\left| \psi \right\rangle $ in
what she decoded with the reliability $100\%$, if what she decoded
is $q$, i. e., she has chosen $c=0$, she must have found $\Phi
^{-}$ or $\Psi ^{+}$ in her measurement. From this equation, we
can see that the system $\beta $ must have collapsed into
\begin{equation}
\left| B_{0}^{\prime }\right\rangle \equiv
\sum_{r}(-1)^{r}f_{r,+}\left| B_{r,+}\right\rangle
/\sqrt{\sum_{r}f_{r,+}^{2}}
\end{equation}
or
\begin{equation}
\left| B_{1}^{\prime }\right\rangle \equiv
\sum_{r}(-1)^{r}f_{r,\times }\left| B_{r,\times }\right\rangle
/\sqrt{\sum_{r}f_{r,\times }^{2}}.
\end{equation}
Similarly, if $c=1$, $\beta $ must have collapsed into
\begin{equation}
\left| B_{0}^{\prime \prime }\right\rangle \equiv
\sum_{q}f_{0,q}\left| B_{0,q}\right\rangle
/\sqrt{\sum_{q}f_{0,q}^{2}}
\end{equation}
or
\begin{equation}
\left| B_{1}^{\prime \prime }\right\rangle \equiv
\sum_{q}f_{1,q}\left| B_{1,q}\right\rangle
/\sqrt{\sum_{q}f_{1,q}^{2}}.
\end{equation}
Therefore if Bob can distinguish $\{\left| B_{k}^{\prime
}\right\rangle \}$ from $\{\left| B_{k}^{\prime \prime
}\right\rangle \}$, he knows Alice's choice of $c$. Define
\begin{equation}
\rho _{0}\equiv \sum_{k}\left| B_{k}^{\prime }\right\rangle \left\langle
B_{k}^{\prime }\right| /2
\end{equation}
and
\begin{equation}
\rho _{1}\equiv \sum_{k}\left| B_{k}^{\prime \prime }\right\rangle
\left\langle B_{k}^{\prime \prime }\right| /2.
\end{equation}
The upper bound (Holevo bound) of the average information Bob can get is
\begin{equation}
I_{av}=S[(\rho _{0}+\rho _{1})/2]-[S(\rho _{0})+S(\rho _{1})]/2,
\end{equation}
where the von Neumann entropy is $S(\rho )=-Tr(\rho \log _{2}\rho )$ \cite
{Holevo}. From the symmetry of the equation, it can be seen that $I_{av}$\
will go to its extremum when Bob chooses $f_{r,q}=1/2$ for all $r$, $q$. It
is found that this extremum is the maximum. That is, if Bob prepares the
initial state as
\begin{equation}
\left| \psi \right\rangle =\sum_{r\in \{0,1\},q\in \{+,\times \}}\left|
r\right\rangle _{q}\left| r\right\rangle _{q}\left| B_{r,q}\right\rangle /2,
\end{equation}
the probability for him to get $c$ with the reliability $100\%$ will be
maximized. In the previous paragraph, it is shown that all $\left|
B_{r,q}\right\rangle $ need to be orthogonal. For illustration, it is
natural to choose
\begin{equation}
\left| B_{r,q}\right\rangle =\left| Q\right\rangle _{+}\left| r\right\rangle
_{+}
\end{equation}
where $Q=0,1$ for $q=+,\times $. Then
\begin{equation}
\left| B_{k}^{\prime }\right\rangle =\left| k\right\rangle _{+}\left|
1\right\rangle _{\times },\qquad \left| B_{k}^{\prime \prime }\right\rangle
=\left| 0\right\rangle _{\times }\left| k\right\rangle _{+}\qquad (k=0,1).
\end{equation}
We can see that $\rho _{0}$ and $\rho _{1}$ are diagonalized simultaneously
in the basis that Bob uses in the projection in the step (4). Therefore this
projection is just the optimal strategy for Bob to decode $c$, and the
maximum probability for the decoding to be successful is $1/2$. If Bob does
not prepare the initial state in this way, this maximum probability cannot
be reached as $I_{av}$\ is not optimized. Similar to the proof in (II), the
probability for him to pass steps (5) can be made arbitrarily small as $%
m\rightarrow \infty $. %This finishes the proof.

Combining points (I)-(III), we can conclude that the probability for Alice
to know whether Bob gets $b$ or not (or the probability for Bob to get $b$
in more than $50\%$ of the cases) is expressed as $O(\delta )\varepsilon
^{O(\delta n)}$ (or $O(\delta )\varepsilon ^{O(\delta m)}$), which is
arbitrarily small by increasing $n$, $m$. Also, unlike the cheat sensitive
protocols\cite{cheat sensitive}, the detection of cheating in our protocol
will not cause the secret bit of OT to be revealed. As a result, the present
Protocol OT is unconditionally secure. As our proof is based on the density
matrices of the quantum states, rather than on a specific cheating strategy,
our conclusion is general no matter what computational power the
participants may have and what POVMs they may apply.

\section{Relationship with the no-go theorems}

\subsection{The Lo's no-go theorem of quantum secure computations}

Though the above general proof of security against all possible cheating
strategies seems complicated, the reason why this protocol can evade the
cheating in the Lo's no-go theorem is clear. As mentioned in the
introduction, the protocol does not satisfy the requirement (a) (Bob learns
a prescribed function $f(i,j)$ \textit{unambiguously}) in Ref.\cite
{impossible1}, on which the no-go proof is based. This is because Bob cannot
learn the value of $b$ \textit{unambiguously} in our protocol. Instead, he
only learns $b$ with the probability $50\%$. In the other $50\%$ case, he
has zero knowledge on $b$. In addition, rigorously speaking, the outcome of
our protocol cannot be viewed as a prescribed function $f(i,j)$. The outcome
depends not only on Alice's and Bob's inputs $i$ and $j$, but also on the
quantum uncertainty in the measurement. For example, in the step (4) of our
protocol, Bob's inputting $d=1$ does not mean that he can certainly obtain
the value of $c$. Due to the quantum uncertainty in his measurement, he can
only obtain $c$ successfully with the probability $50\%$.
%Whether Bob finally gets $b$ or not is
%undoubtedly affected by this uncertainty.
As a result, the quantum state in our protocol is not the simultaneous
eigenstate of different measurement operators that the participant uses for
determining the parameters wanted by him (e.g., $s$ and $c$). He knows
whether he gets a parameter successfully only if the measurement is
performed. Then the state is disturbed, so that it cannot be used to get
more parameters. Thus the protocol is secure against the cheating strategy
in Ref.\cite{impossible1}. On the other hand, the definition of
all-or-nothing OT only requires that at the end of the protocol, the two
outcomes ``Bob learns the value of $b$'' and ``Bob has zero knowledge on $b$%
'' should occur with the equal probability $50\%$; while it never requires
that which outcome finally happens must be controlled only by the
participants' inputs. Clearly, our protocol satisfies the rigorous
definition of secure all-or-nothing OT.

\subsection{The MLC no-go theorem of secure QBC}

Our result does not conflict with the MLC no-go theorem of secure QBC,
because this no-go theorem does not apply directly to QOT (otherwise the
Lo's no-go theorem of quantum secure computation would be redundant). Let $%
P_{1}$ denote an all-or-nothing QOT protocol. Surely it does not implement
QBC automatically. Instead, another protocol $P_{2}$ is needed, which makes
use of the output of $P_{1}$ to accomplish QBC. The MLC no-go theorem
reveals that the entire protocol $P_{1}+P_{2}$ cannot be secure. Then there
are two possibilities: $P_{1}$ is insecure, or $P_{2}$ is insecure (if not
both). But as we already proved rigorously in Sec. III, our all-or-nothing
QOT protocol is unconditionally secure against any cheating strategy.
Therefore the existence of the MLC no-go theorem implies that secure $P_{2}$
is impossible.

Indeed, though BC and OT are thought to be classically equivalent,
``reductions and relations between classical cryptographic tasks need not
necessarily apply to their quantum equivalents''\cite{string BC}. So far
there are two known methods to construct $P_{2}$\ in classical cryptography,
which all fail at the quantum level. One of the method is to repeat
all-or-nothing OT many times\cite{Kilian}. More rigorously, according to
Ref. \cite{Kilian}, BC is realized by encoding the committed bit as $%
b=b_{1}\oplus b_{2}\oplus ...\oplus b_{k}$, and sending each $b_{i}$ from
Alice to Bob through an all-or-nothing OT process. However, the resultant
protocol is insecure because altering anyone of the $b_{i}$ can flip the
value of the committed bit completely. Alice can simply execute the protocol
honestly. If she wants to change the committed bit at the final stage, she
simply announces one of the $b_{i}$ dishonestly. Since Bob knows $b_{i}$\ at
half of the cases\ only, Alice can cheat successfully with the probability $%
1/2$. Thus the scheme is broken. Another known method to realize BC from OT
in classical cryptography is to build an 1-out-of-2 OT\cite{p-OT,1-2OT} from
all-or-nothing OT, and use the 1-out-of-2 OT to implement BC. But once
again, it has to rely on the classical equivalence between 1-out-of-2 OT and
all-or-nothing OT\cite{p-OT}, which needs re-examination at the quantum
level. As pointed out in Ref.\cite{impossible1}, classical reduction would
be applicable in quantum cryptography if a quantum protocol can be used as a
''black box'' primitive in building up more sophisticated protocols.
However, we found recently\cite{2OT} that the 1-out-of-2 OT protocol built
upon the present quantum all-or-nothing OT protocol with the scenario
developed in Ref.\cite{p-OT} is not rigorously a ''black box'' type quantum
1-out-of-2 OT specified in Ref.\cite{impossible1}. Especially, the inputs of
the two participants are not independent of each other. Such a quantum
1-out-of-2 OT cannot be used to implement secure QBC with the method
described in Ref.\cite{Short}. The reason lies in that the step (2) of the
protocol described in Ref.\cite{Short} is inexecutable as Alice's input
cannot be completed before Bob's input is entered. Thus the method also
fails. Of course there may exist other methods to construct $P_{2}$, but due
to the presence of the MLC no-go theorem, they are all bound to be insecure.
In this sense, the classical reduction chain from OT to BC is broken in the
present quantum case, and thus there exists no logic conflict between the
present secure all-or-nothing QOT and the MLC no-go theorem of QBC.

\section{Summary and Discussions}

In all, we proposed an quantum all-or-nothing oblivious transfer protocol
based on quantum entangled states, and proved that it is unconditionally
secure against any cheating strategy. It was also illustrated how the
protocol evades the Lo's no-go theorem of the one-sided two-party secure
computation, as well as that the security of our QOT does not conflict with
the MLC no-go theorem of QBC.

The existence of secure QOT protocol is important not only for
multi-party protocols, but also for a better understanding of
quantum theory. According to recent results\cite{philosophy},
three fundamental information-theoretic constrains, namely, the
impossibilities of (i) superluminal information transfer between
two physical systems by performing measurements on one of them;
(ii) broadcasting the information contained in  unknown physical
states; and (iii) unconditionally secure bit commitment, may
suffice to entail that the observables and state space of a
physical theory are quantum-mechanical. Therefore, clarifying the
boundary between the capability and limitation of quantum
cryptography, as well as the relationship between classical
cryptography and its quantum counterpart, can certainly enrich our
knowledge for  searching  the answer to Wheeler's query ``Why the
quantum''\cite {philosophy}.

Finally, it is worth pinpointing that a QBC protocol somewhat similar to
ours was proposed\cite{Shimizu}: both protocols start with a 4-level system
on Alice's side and rely on a verification procedure to avoid cheating .
However, as pointed out by the authors, what they achieved in Ref.\cite
{Shimizu} was merely an analog to OT, which does not meet the rigorous
security requirement of the OT definition; in fact, they merely attempted to
use the analog to realize a QBC protocol.
% with the completeness that is achieved for the BB84
%protocol for QKD''. %Though bit commitment implies OT in classical
%cryptography, there was also evidence that reductions and
%relations between classical cryptography sometimes do not apply to
%their quantum equivalents\cite{string BC}. Moreover, a single
%QBC-based OT process needs to execute the QBC protocol many
%times\cite{Yao}, which will be much inconvenient and involves tons
%of qubits. On the contrary,
In contrast, our protocol includes a further crucial verification on Bob's
side, %which uses the entanglement to fulfil quantum
%computation and thus surpass theirs,
possessing at least three advantages: (i) the strict requirement of OT is
met; (ii) the stand-alone security is proven to be unconditional; and (iii)
it is convenient to modify ours to be a $p$-OT protocol\cite{p-OT}.
%Besides, the idea of fully
%utilizing of the quantum entanglement to run quantum algorithm in
%protocols is a novelty in quantum cryptography. It may have more
%other applications in cryptography in the forthcoming future,
%which may help to develop the potential of quantum cryptography
%than what it is currently realized.

We thank Hoi-Fung Chau and Hoi-Kwong Lo for their useful
discussions. The work was supported by the RGC grant of Hong Kong
(HKU7114/02P and HKU7045/05P).


\begin{thebibliography}{99}
\bibitem{Ra81}  M. Rabin, technical report TR-81, Aiken Computation
Laboratory, Harvard University, 1981

\bibitem{Wiesner}  S. Wiesner, SIGACT News, \textbf{15}, 78 (1983).

\bibitem{Kilian}  J. Kilian, in \textit{Proceedings of 1988 ACM Annual
Symposium on Theory of Computing}, May 1988, pp.20 (ACM, New York, 1988).

\bibitem{Crepeau95}  C. Cr\'{e}peau, J. van de Graaf, and A. Tapp, in
\textit{Advances in Cryptology: Proceedings of Crypto '95}, Vol.\textbf{963}%
, pp.110 (Springer-Verlag, Berlin, 1995).

\bibitem{Shor}  P. W. Shor, in \textit{Proceedings of the 35th Annual
Symposium on the Foundations of Computer Science}, pp.124 (IEEE Computer
Society, Los Alamitos, CA, 1994).

\bibitem{BB84}  C. H. Bennett, and G. Brassard, in \textit{Proceedings of
IEEE International Conference on Computers, Systems, and Signal Processing},
Bangalore, India, pp.175 (IEEE, New York, 1984).

\bibitem{Ekert}  A. K. Ekert, Phys. Rev. Lett. \textbf{67}, 661(1991); C. H.
Bennett, \textit{ibid.} \textbf{68}, 3121 (1992); D. A. Meyer, \textit{ibid.}
\textbf{82}, 1052 (1999).

\bibitem{BBCS92}  C. Cr\'{e}peau and J. Kilian, in \textit{29th Symp. on
Found. of Computer Sci.}, pp. 42-52 (IEEE, 1988). C. H. Bennett, G.
Brassard, C. Cr\'{e}peau, and M. -H. Skubiszewska, in \textit{Advances in
Cryptology: Proceedings of Crypto '91, }Vol.\textbf{576}, pp.351
(Springer-Verlag, 1992).

\bibitem{OT}  C. Cr\'{e}peau, Journal of Modern Optics,
%special issue on Quantum Communication and Cryptography.
\textbf{41}, 2445 (1994).

\bibitem{Yao}  A. C. C. Yao, in \textit{Proceedings of the 26th Symposium on
the Theory of Computing}, 1995, pp.67 (ACM, New York, 1995).

\bibitem{BCJL93}  G. Brassard, C. Crepeau, R. Jozsa, and D. Langlois, in
\textit{Proceedings of the 34th Annual IEEE Symposium on Foundations of
Computer Science, 1993}, pp.362 (IEEE, Los Alamitos, 1993).

\bibitem{Mayers}  D. Mayers, Phys. Rev. Lett. \textbf{78}, 3414 (1997);
H.-K. Lo and H. F. Chau, Phys. Rev. Lett.\textbf{78}, 3410 (1997).

\bibitem{Lo}  G. Brassard, C. Cr\'{e}peau, D. Mayers, L. Salvail,
quant-ph/9712023.

\bibitem{CT1}  D. A. Meyer, Phys. Rev. Lett. \textbf{82}, 1052 (1999).

\bibitem{CT2}  S. J. van Enk, Phys. Rev. Lett. \textbf{84}, 789 (2000).

\bibitem{CT3}  L. Goldenberg, L. Vaidman, and S. Wiesner, Phys. Rev. Lett.
\textbf{82}, 3356 (1999).

\bibitem{OMI}  C. Crepeau, and L. Salvail, in \textit{Advances in
Cryptology: Proceedings of Eurocrypt '95}, pp.133 (Springer-Verlag, 1995).

\bibitem{MS94}  D. Mayers, and L. Salvail, in \textit{Proceedings of the
Third Workshop on Physics and Computation-PhysComp '94}, pp.69 (IEEE
Computer Society Press, Dallas, 1994).

\bibitem{impossible1}  H.-K. Lo, Phys. Rev. A, \textbf{56}, 1154 (1997).

\bibitem{impossible2}  H. F. Chau and H.-K. Lo, Fortsch. Phys. \textbf{46},
507 (1998).

\bibitem{note}  Here as well as in the steps (3-4), (5-1) and (5-3), a small
deviation from the mean value of the equidistribution is tolerable. The
range of tolerable deviation can be evaluated by the Bernshtein's law of
large numbers %\cite{BCJL93}
and then agreed on by the participants. It can be proven that by
choosing sufficiently large security numbers $n$, $m$ and the
number of $\left| \psi \right\rangle $ picked for the
verifications, the protocol can be successful with a probability
larger than $1-x$, while the probability of the cheating
\textit{via} this deviation is less than $x$. Here $x$ can be made
arbitrarily small.

\bibitem{qi73}  L. P. Hughston, R. Jozsa, and W. K. Wootters, Phys. Lett. A
\textbf{183}, 14 (1993).

\bibitem{Holevo}  A. S. Holevo, \textit{Probabilistic and Statistical
Aspects of Quantum Theory} (North-Holland, Amsterdam, 1982).

\bibitem{cheat sensitive}  L. Hardy and A. Kent, Phys. Rev. Lett. \textbf{92}%
, 157901 (2004).

\bibitem{string BC}  A. Kent, Phys. Rev. Lett. \textbf{90}, 237901 (2003).

\bibitem{p-OT}  C. Cr\'{e}peau, in \textit{Advances in Cryptology:
Proceedings of Crypto '87}, Vol.\textbf{293}, pp. 350 (Springer-Verlag,
1988).

\bibitem{1-2OT}  S. Even, O. Goldreich, and A. Lempel, \textit{Advances in
Cryptology: Proceedings of Crypto '82}, pp.205 (Plenum 1982).

\bibitem{2OT}  G. P. He and Z. D. Wang, quant-ph/0504170.

\bibitem{Short}  T. Short, N. Gisin, and S. Popescu, quant-ph/0504134.

\bibitem{philosophy}  R. Clifton, J. Bub and H. Halvorson, Found. Phys.
\textbf{33}, 1561 (2003). H. Halvorson, J. Bub, quant-ph/0311065.

\bibitem{Shimizu}  K. Shimizu, and N. Imoto, Phys. Rev. A \textbf{66},
052316 (2002); Phys. Rev. A \textbf{67}, 034301 (2003).

%\bibitem{full}  G. P. He and Z. D. Wang, quant-ph/0312161(v.2).

%\bibitem{coin}  A.Ambainis, e-print no. quant-ph/0204063, (2002).

\end{thebibliography}
\end{document}